\documentclass[aps,pra,showpacs,floatfix,superscriptaddress,twocolumn]{revtex4}
\usepackage[dvips]{graphicx}
\usepackage{color}

\usepackage{longtable}
\usepackage{dcolumn}
\usepackage[dvips]{graphicx}
\usepackage{bm}
\usepackage{bbm}

\usepackage{times}
\usepackage{nicefrac}
\usepackage{amsmath}
\usepackage{amsfonts}
\usepackage{amssymb}
\usepackage{amsthm}

\newcolumntype{.}{D{x}{}{-1}}

\newcommand{\balpha}{\bm{\alpha}}

\newcommand{\vare}{\varepsilon}
\newcommand{\bfr}{{\bm {r}}}

\newcommand{\bfp}{{\bm {p}}}
\newcommand{\bfq}{{\bm {q}}}

\newcommand{\lbr}{\langle}
\newcommand{\rbr}{\rangle}

\newcommand{\pr}{^{\prime}}

\begin{document}

\title{Target effects in negative-continuum assisted dielectronic recombination}

\author{V. A. Yerokhin}

\affiliation{Center for Advanced Studies, Peter the Great St.~Petersburg Polytechnic University,
Polytekhnicheskaya 29, 195251 St.~Petersburg, Russia}

\affiliation{Helmholtz-Institut Jena, Fr\"obelstieg 3, D-07743 Jena, Germany}

\author{A. N. Artemyev}

\affiliation{Institut f\"ur Physik, Universit\"at Kassel, Heinrich-Plett-Str. 40, 34132, Kassel,
Germany}

\author{V. M. Shabaev} \affiliation{Department of Physics, St.~Petersburg State University, Ulianovskaya 1, Petrodvorets,
St.Petersburg 198504, Russia}

\author{Th. St\"ohlker}

\affiliation{Helmholtz-Institut Jena, Fr\"obelstieg 3, D-07743 Jena, Germany} \affiliation{Institut
f\"ur Optik und Quantenelektronik, Friedrich-Schiller-Universit\"at Jena, 07743 Jena, Germany}

\author{A. Surzhykov}

\affiliation{Helmholtz-Institut Jena, Fr\"obelstieg 3, D-07743 Jena, Germany}

\author{S. Fritzsche}

\affiliation{Helmholtz-Institut Jena, Fr\"obelstieg 3, D-07743 Jena, Germany}
\affiliation{Theoretisch-Physikalisches Institut, Friedrich-Schiller-Universit\"at Jena,
F\"urstengraben 1, 07743 Jena, Germany}

\begin{abstract}

The process of recombination of a quasi-free electron into a bound state of an initially bare
nucleus with the simultaneous creation of a bound-electron--free-positron pair is investigated.
This process is called the negative-continuum assisted dielectronic recombination (NCDR). In a
typical experimental setup, the initial electron is not free but bound in a light atomic target.
In the present work, we study the effects of the atomic target on the single and
double-differential cross sections of the positron production in the NCDR process. The
calculations are performed within the relativistic framework based on QED theory, with accounting
for the electron-electron interaction to first order in perturbation theory. We demonstrate how
the momentum distribution of the target electrons removes the non-physical singularity of the
differential cross section which occurs for the initially free and monochromatic electrons.

\end{abstract}


\maketitle

\section{Introduction}

One of the main processes that occurs in the collision of an electron with a bare nucleus is the
radiative recombination of the electron into a bound state under the simultaneous emission of a
photon \cite{eichler:95:book,eichler:07:review}. If the energy of the initial electron is high
enough, new recombination channels become energetically possible. In the present work we are
interested in one of these channels in which the recombination of the initial electron leads to the
creation of an electron-positron pair. The electron from this pair is captured into a bound state
of the ion, whereas the positron is emitted into the continuum. Such process is called the {\em
negative-continuum assisted dielectronic recombination} (NCDR)
\cite{artemyev:03:ncdr,artemyev:09:ncdr},
\begin{align}
X^{Z+} + e^- \to X^{(Z-2)+} + e^+\,.
\end{align}
We refer here to the analogy between NCDR and the well-known dielectronic recombination (DR)
\cite{bell:85,hahn:88,spies:92,brandau:03,brandau:08}. The difference between the two processes is
that in DR the second electron is excited from an initially bound state, whereas in NCDR it is
``excited" from the negative continuum.

In contrast to the dielectronic recombination, NCDR is a nonresonant process, with the energy
threshold condition
\begin{align}
T_i \ge 2\,mc^2 -E_{\rm io}(a)-E_{\rm io}(b)\,,
\end{align}
where $T_i$ is the kinetic energy of the initial-state electron in the nucleus rest frame, $m$ is
the electron mass, and $E_{\rm io}(a)$ and $E_{\rm io}(b)$ are the ionization energies of the first
and second electron in the final state, respectively. For the bare uranium nucleus, the threshold
energy is $T_i \approx 760$~keV.

The NCDR process has a distinct signature that facilitates its experimental identification, namely,
the coincidence of the emitted positron with the doubly charge-exchanged ion. Other possible
(two-step) processes that result in the same final state of the ion and positron were shown
\cite{najjari:15} to yield much smaller contributions, well below of what can be resolved under the
present and near-future experimental conditions. The NCDR has not been observed until now but
should become accessible at the High-Energy Storage Ring (HESR) facility at the future FAIR
acceleration complex. Indeed, such experiments are planned \cite{FAIR}; a first feasibility study
has been presented in Ref.~\cite{hillenbrand:15}.

Theoretical investigations of the NCDR process have been carried out in
Refs.~\cite{artemyev:03:ncdr,artemyev:09:ncdr} under the assumption that the initial-state nucleus
$X^{Z+}$ is colliding with the monochromatic electron beam. The presently prepared experiments,
however, are going to use a somewhat different scenario \cite{hillenbrand:15}. In it, a heavy
energetic bare nucleus will collide with a light atomic target $A$ (which is at rest in the
laboratory frame). For appropriate collision energies, the nucleus can capture one electron from
the target, while another electron is excited from the negative continuum, producing a positron in
continuum,
\begin{align}
X^{Z+} + A \to X^{(Z-2)+} + A^{+} + e^+\,.
\end{align}
Here we aim to analyze the NCDR process within this scenario that may be realized in a future
experimental setup at the HESR.

In practice, the atomic target affects the NCDR process in a two-fold manner. Firstly, it
determines the shape of the energy spectra of the emitted positron; secondly, it removes the
nonphysical singularity of the angle-differential NCDR cross section in the laboratory frame as
reported in Refs.~\cite{artemyev:03:ncdr,artemyev:09:ncdr}. In the present work we shall consider
both these effects in detail.

The paper is organized as follows. Section~\ref{sec:1} describes the theory of the NCDR process
within the relativistic framework. Numerical details are discussed in Sec.~\ref{sec:2}.
Sec.~\ref{sec:3} presents the results and discussion. Finally, a short summary is given in
Sec.~\ref{sec:4}. Relativistic units $\hbar=c=1$ are used throughout the paper.

\section{Theory}
\label{sec:1}

\subsection{Monochromatic initial-state electrons}
\label{sec:01}

We start from the same scenario of monochromatic initial-state electrons as was considered
previously in Refs.~\cite{artemyev:03:ncdr,artemyev:09:ncdr}. In the rest frame of the nucleus, the
incoming electron possesses the asymptotic momentum $\bfp_i$, the energy $\vare_i =
(\bfp_i^2+m^2)^{1/2}$, and the helicity $m_i = \pm 1/2$. In the final state we have a continuum
positron with the asymptotic momentum $\bfp_f$, energy $\vare_f = (\bfp_f^2+m^2)^{1/2}$, and the
momentum projection $m_f$ as well as the two-electron bound state $|n_a\kappa_a,n_b\kappa_b;JM\rbr$
with the total angular momentum $J$ and projection $M$. Here, $n_a$, $\kappa_a$ and $n_b$,
$\kappa_b$ denote the principal quantum number and the relativistic angular quantum number of the
electrons states $a$ and $b$, respectively.

\begin{figure}
\centerline{
\resizebox{0.5\columnwidth}{!}{%
  \includegraphics{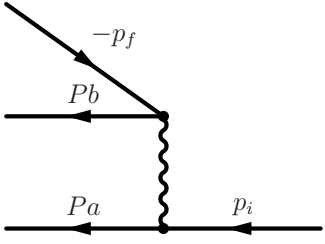}
}}
 \caption{
Schematic representation of the negative-continuum assisted dielectronic recombination. $p_i$ denotes the
incoming quasi-free electron, $-p_f$ represents the outgoing positron, $a$ and $b$ are bound
one-electron states, and $P$ is the permutation operator, $PaPb = ab$ or $ba$. \label{fig1}}
\end{figure}

Following the general QED theory \cite{shabaev:02:rep} (see also
Refs.~\cite{artemyev:03:ncdr,artemyev:09:ncdr}),  the angle-differential cross section of the NCDR
process in the nucleus-rest frame is given by
\begin{align} \label{eq:01}
\frac{d \sigma}{d \Omega_f} &\ = \frac{16\pi^4N^2}{v_i}\,\vare_f\,|\bfp_f|\, \sum_{m_i m_f M}
\biggl| \sum_{m_am_b}C^{JM}_{j_am_a,j_bm_b}\,
 \nonumber \\ & \times
\sum_P(-1)^P\,
\lbr PaPb|I(\vare_i-\vare_{Pa})|p_im_i,-p_fm_f\rbr\biggr|^2\,,
\end{align}
where $v_i$ is the speed of the incoming electron, $P$ is the permutation operator, $PaPb = (ab)$
or $(ba)$, $(-1)^P$ is the sign of the permutation, $|ab\rbr \equiv
|n_a\kappa_am_a,n_b\kappa_bm_b\rbr$; $N = 1/2$ when $n_a = n_b$ and $\kappa_a = \kappa_b$ and $N =
1/\sqrt{2}$ otherwise; $\vare_a$ and $\vare_b$ are the one-electron energies of the electron states
$a$ and $b$, respectively; $|pm\rbr$ denotes the electron continuum Dirac state with a definite
asymptotic 4-momentum $p \equiv (\vare,\bfp)$ and a momentum projection $m$, and $I(\omega)$ is the
relativistic operator of the electron-electron interaction. In the Feynman gauge, $I(\omega)$ reads
\begin{align}
I(\omega) = \alpha\,(1-\balpha_1\cdot\balpha_2)\,\frac{\exp(i|\omega|r_{12})}{r_{12}}\,,
\end{align}
where $\alpha$ denotes the fine-structure constant, $r_{12} = |\bfr_1-\bfr_2|$ is the distance
between the electrons, and $\balpha_k$ denotes the vector of Dirac matrices acting on the $k$th
electron.

In Eq.~(\ref{eq:01}) we have assumed  that the incident electron is unpolarized and that the spin
states of the emitted positron as well as the momentum projections of the residual ion remain
unobserved. Following the standard procedure \cite{itzykson:80,bjorken:64}, we have replaced the
positron with four-momentum $p_f$ and momentum projection $m_f$ by an electron with four-momentum
$-p_f$ and momentum projection $-m_f$. Fig.~\ref{fig1} displays this process schematically in terms
of its Feynman diagram.

Eq.~(\ref{eq:01}) is exact in leading order of QED perturbation theory with regard to the
electron-electron interaction. Higher-order electron-electron interaction effects are suppressed
typically by a parameter of $1/Z$, where $Z$ is the nuclear charge of the projectile. For heavy
projectiles, as those considered in the present work, these effects are small and negligible.

The energy of the emitted positron in Eq.~(\ref{eq:01}) is fixed by the energy conservation
condition (in the rest frame of the nucleus),
\begin{align}
\vare_f = \vare_i - \vare_a - \vare_b\,,
\end{align}
or
\begin{align}
T_f = T_i - 2m + E_{\rm io}(a) + E_{\rm io}(b)\,,
\end{align}
where $E_{\rm io}(a)$ and $E_{\rm io}(b)$ are the ionization energies of the electrons $a$ and $b$,
respectively. Within the approximation of Eq.~(\ref{eq:01}), the final-state electrons $a$ and $b$
are considered within the independent particle model, so the energy of the final bound state of the
helium-like ion is given by the sum of the Dirac energies of the electrons $a$ and $b$.

It should be noted that Eq.~(\ref{eq:01}) differs from the analogue formulas in
Refs.~\cite{artemyev:03:ncdr,artemyev:09:ncdr} by the prefactor  $\vare_f/|\bfp_f|$. This prefactor
was omitted in the previous studies due to a mistake in the derivation. Because of this, all
numerical results for the cross sections in Refs.~\cite{artemyev:03:ncdr,artemyev:09:ncdr} should
be multiplied by $\vare_f/|\bfp_f|$. The numerical value of this prefactor is different for
different initial electron energies as well as different nuclear charges. In the particular case of
uranium and the incoming electron energy $T_i = 1300$~keV, the additional prefactor amounts to
$\approx 1.4$.

The evaluation of the matrix elements of the operator $I(\omega)$ in Eq.~(\ref{eq:01}) has been
discussed in detail in Refs.~\cite{artemyev:03:ncdr,artemyev:09:ncdr}, so it need not be repeated
here.

\subsection{Effects of atomic target} \label{sec:02}

In Sec.~\ref{sec:01} we studied the NCDR process in the rest frame of the nucleus, assuming that
the incoming electron is free and has a definite energy and asymptotic momentum. We now consider a
more realistic scenario, in which the initial-state electron is bound in a light atomic target. In
the laboratory frame, the atomic target is at rest, whereas the bare nucleus (projectile) is moving
with the reduced velocity $\beta = v/c$ and the Lorentz factor $\gamma = (1-\beta^2)^{-1/2}$ along
the $z$ axis.

We shall assume that the projectile velocity $\beta$ is much larger than the typical value of
$|\bfq|/m$, $\bfq$ being the momentum of  a target electron with respect to the target nucleus.
Under this assumption, the so-called impulse approximation \cite{kleber:75} is valid, which
describes the atomic target as a collection of independent free electrons with a momentum
distribution determined by the bound electron orbitals in the target.

From now on, we shall distinguish between the laboratory frame (unprimed variables) and the
projectile frame, i.e., the nucleus rest frame (primed variables).

Within the impulse approximation, the double differential cross section of the NCDR process with
capture from the atomic state $|i\rbr \equiv |njlm\rbr$ can be written as [see
Ref.~\cite{eichler:07:review}, Eq. (3.4)] as
\begin{align} \label{eq:10}
\frac{d^2\sigma\pr}{d\Omega_f\pr\,d\vare_f\pr} =  \frac1{2j+1}\sum_m \int d\bfq\, \bigl|  &\ \psi_{njlm}(\bfq)\bigr|^2\,
 \frac{d\sigma(p_i\pr)}{d\Omega_f\pr}\,
  \nonumber \\ & \times
 \delta(\vare_f\pr + \vare_a\pr + \vare_b\pr - \vare_i\pr)\,,
\end{align}
where $\psi_{njlm}(\bfq)$ is the momentum representation of the target electron wave function (in
laboratory frame) and $d\sigma(p_i\pr)/d\Omega_f\pr$ is the single differential NCDR cross section
as given by Eq.~(\ref{eq:01}). The $\delta$-function expresses the energy conservation in the
projectile frame between the initial-state electron energy $\vare_i\pr$ and the emitted positron
energy $\vare_f\pr$.

In the laboratory frame, the initial-state electron has the energy $\vare_i \equiv m - E_{\rm
io}(i)$ and momentum $\bfq$. In the projectile system, the energy of the electron is $\vare_i\pr =
\gamma \vare_i - \gamma v q_z$, where $q_z$ is the projection of $\bfq$ on the direction of the
projectile propagation. We  can thus rewrite the $\delta$ function in Eq.~(\ref{eq:10}) as
\begin{align} \label{eq:11}
\delta(\vare_f\pr + \vare_a\pr + \vare_b\pr - \vare_i\pr) =
\delta\bigl(\vare_f\pr - \vare_{f,0}\pr + \gamma E_{\rm io}(i) + \gamma v q_z\bigr)\,,
\end{align}
where $\vare_{f,0}\pr = \gamma mc^2 - \vare_a\pr - \vare_b\pr$. Integration over $q_z$ in
Eq.~(\ref{eq:10}) can be performed with help of the $\delta$ function. The result is
\begin{align} \label{eq:12}
\frac{d^2\sigma\pr}{d\Omega_f\pr\,d\vare_f\pr} = \frac1{2j+1}\sum_m \frac1{\gamma v} \int d^2\bfq_{\perp}\,
    \bigl| \psi_{njlm}(\bfq)\bigr|^2\,
 \frac{d\sigma(p_i\pr)}{d\Omega_f\pr}\,,
\end{align}
where the integration is performed over the transverse momentum $\bfq_{\perp}$, whereas the value
of the longitudinal momentum is fixed by $ q_z = [\vare_{f,0}\pr -\vare_f\pr - \gamma E_{\rm
io}(i)]/\gamma v$.

Taking into account that the momentum distribution of the target electrons is peaked around $\bfq
\approx 0$ and that the cross section only weakly depends on the direction of $\bfq$, we can move
the cross section outside the integral,
\begin{align} \label{eq:12}
\frac{d^2\sigma\pr}{d\Omega_f\pr\,d\vare_f\pr} = \frac1{\gamma v}
   \frac{d\sigma(p_i\pr)}{d\Omega_f\pr}\, {\cal L}_{njl}(q_z)
\,,
\end{align}
where
\begin{align} \label{eq:13}
{\cal L}_{njl}(q_z) =
 \frac1{2j+1}\sum_m
\int d^2\bfq_{\perp}\,
    \bigl| \psi_{njlm}(\bfq)\bigr|^2\,
\end{align}
is the Compton profile of the target electron state $|njl\rbr$. We now take into account that
$d^2\bfq_{\perp} = q_{\perp} dq_{\perp} d\varphi = q\, dq\, d\varphi$, where $q = |\bfq| =
\sqrt{q_z^2 + \bfq^2_{\perp}}$. The integration over the polar angle in Eq.~(\ref{eq:13}) is
performed by using the identity
\begin{align} \label{eq:14}
 \frac1{2j+1}\sum_m
    \bigl| \psi_{njlm}(\bfq)\bigr|^2 = \frac1{4\pi}\,\bigl[ g_{njl}^2(q) + f_{njl}^2(q)\bigr]\,,
\end{align}
where $g_{njl}(q)$ and $f_{njl}(q)$ are the upper and the lower radial wave-function components in
the momentum representation. In the result, we have the Compton profile of the electron state
$|njl\rbr$
\begin{align} \label{eq:15}
{\cal L}_{njl}(q_z) =
\frac12\,\int_{q_z}^{\infty}q\,dq\, \bigl[ g_{njl}^2(q) + f_{njl}^2(q)\bigr]\,,
\end{align}
which is normalized as
\begin{align} \label{eq:16}
2\,\int_0^{\infty}dq_z\,{\cal L}_{njl}(q_z) =1\,.
\end{align}

Finally, since we are interested in the cross section of the NCDR process, independent of the
atomic shell from which the electron is captured, we need to sum up over all occupied atomic shells
of the target,
\begin{align} \label{eq:17}
\frac{d^2\sigma\pr}{d\Omega_f\pr\,d\vare_f\pr} = \frac1{\gamma v}\, \sum_{a}g_{a}\,
   \frac{d\sigma(p_i\pr)}{d\Omega_f\pr}\, {\cal L}_{a}(q_z)
\,,
\end{align}
where $g_{a}$ is the occupation number of the shell $a$.

\subsection{Transformation to the laboratory frame}
\label{sec:Lorentz}

So far, we considered the NCDR cross section in the projectile frame. In order to provide results
that can be interpreted in an experiment, we need to transform our formulas to the laboratory
frame. The well-known Lorenz transformation rules from the projectile frame (primed variables) to
the laboratory frame (unprimed variables) are
\begin{align} \label{eq:20}
p_f\, \sin\theta_f &\ = p_f\pr\,\sin \theta_f\pr\,, \\ \label{eq:20:b}
p_f\, \cos\theta_f &\ = \gamma(p_f\pr\,\cos \theta_f\pr + \beta \vare_f\pr/c)\,,  \\
\vare_f &\ = \gamma (\vare_f\pr + v\,p_f\pr \cos\theta_f\pr)\,,\label{eq:20:c}
\end{align}
where $\theta_f$ is the polar angle with respect to the direction of movement of the projectile
frame.

The transformation of the double-differential cross section (\ref{eq:17}) to the laboratory frame
is given by (see Eq.~(2.49) of Ref.~\cite{eichler:95:book}),
\begin{align} \label{eq:21}
\frac{d^2\sigma}{d\Omega_f\,d\vare_f} = \frac{p_f}{p_f\pr}\, \left( \frac{d^2\sigma\pr}{d\Omega_f\pr\,d\vare_f\pr}
\biggr|_{\theta_f\pr \to \pi-\theta_f\pr}\right) \,,
\end{align}
where $\theta_f\pr \to \pi-\theta_f\pr$ refers to the additional mirror reflection of the angle in
the projectile frame. This reflection is needed since the propagation direction of the projectile
in the laboratory system is opposite to the propagation direction of the incoming electron in the
projectile system.

From Eqs.~(\ref{eq:20}) and (\ref{eq:20:b}) we obtain
\begin{align} \label{eq:22}
\frac{p_f}{p_f\pr} = \frac{\sin \theta_f\pr}{\sin \theta_f} = \bigl[ \gamma^2(g+ \cos\theta_f\pr)^2 + \sin^2 \theta_f\pr\bigr]^{1/2}\,,
\end{align}
where $g = \beta/\beta_f\pr \equiv  \beta \vare_f\pr/(p_f\pr c)$.

\begin{figure}
\centerline{
\resizebox{\columnwidth}{!}{%
  \includegraphics{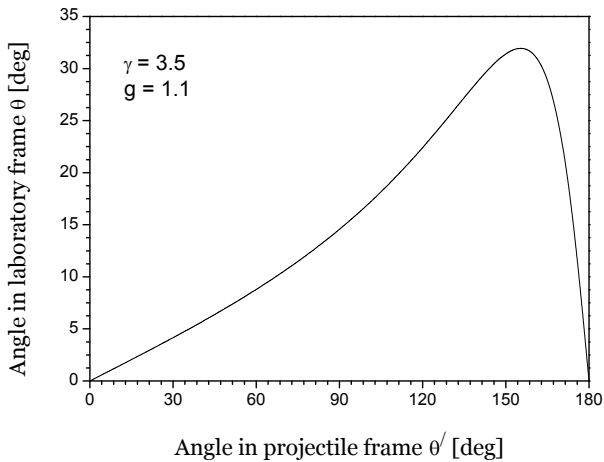}
}}
 \caption{
Transformation of angles between the projectile frame and the laboratory frame, for
collision U$^{92+}+\,$Ar at 2.37~GeV/u,
at the maximum of the positron-energy spectrum.
\label{fig2}}
\end{figure}

An interesting feature of the NCDR process is the absence of the one-to-one correspondence between
the angles in the projectile and in the laboratory frame. The reason is that for NCDR the reduced
velocity of the projectile system $\beta$ is  larger than the reduced velocity of the emitted
positron in the projectile frame $\beta_f\pr$, i.e., $g > 1$. The typical angle transformation plot
in this case is shown in Fig.~\ref{fig2}. From this figure we observe, first, that the positron
emission angle in the laboratory systems cannot exceed some maximal value $\theta_{\rm max}$ with
$\tan\, \theta_{\rm max} = 1/(\gamma\sqrt{g^2-1})$ and, second, that any angle $\theta_f \neq
\theta_{\rm max}$ in the laboratory frame corresponds to {\em two} angles in the projectile frame,
$\theta_{f,1}\pr$ and $\theta_{f,2}\pr$ \cite{dedrick:62}. Moreover, from Eq.~(\ref{eq:20:c}) we
deduce that the positron energy in the laboratory frame $\vare_f$ depends on the emission angle in
the projectile system $\theta_f\pr$. Therefore, every peak of the positron energy spectrum in the
projectile system is transformed into {\em two} peaks in the laboratory frame.

For completeness, we also present the Lorentz transformation of the angle-differential NCDR cross
section (\ref{eq:01}) from the projectile frame into the laboratory frame,
\begin{align} \label{eq:23}
\frac{d\sigma}{d\Omega_f} = \left| \frac{d\,\cos \theta_f\pr}{d\,\cos  \theta_f}\right|\,
\left( \frac{d\sigma\pr}{d\Omega_f\pr}
\biggr|_{\theta_f\pr \to \pi-\theta_f\pr}\right)\,,
\end{align}
where the Jacobian of angle transformation is given by
\begin{align} \label{eq:24}
\left| \frac{d\,\cos \theta_f\pr}{d\,\cos \theta_f} \right| =
 \frac{\bigl[\gamma^2(g+\cos\theta_f\pr)^2+\sin^2 \theta_f\pr \bigr]^{3/2}}{\gamma |1+g\cos \theta_f\pr|}\,.
\end{align}
The transformation (\ref{eq:23}) is encountered if we assume (as in
Refs.~\cite{artemyev:03:ncdr,artemyev:09:ncdr}) the asymptotic momentum of the incoming electron to
be fixed, i.e., if we neglect the momentum distribution of the electrons in the target.

It is remarkable that the angle transformation (\ref{eq:24}) is singular for $g \ge 1$, as the
denominator vanishes at the critical angle $\cos \theta\pr_{\rm max} = -1/g$.  In contrast to that,
the angle transformation for the double differential cross section (\ref{eq:22}) does not contain
any singularities. (The singularity may be recovered if we assume a $\delta$-function energy
distribution of target electrons and integrate over the electron energy.) In the present work we
will demonstrate that the singularity of the cross section in the laboratory frame disappears if we
assume any reasonable momentum distribution of target electrons.

\section{Numerical details}
\label{sec:2}

The evaluation of the angle-differential NCDR cross section ${d\sigma}/{d\Omega_f}$ defined by
Eq.~(\ref{eq:01}) was discussed in detail in Refs.~\cite{artemyev:03:ncdr,artemyev:09:ncdr} and
thus will not be repeated here. In the present work we need to convolute the the angle-differential
cross section with the Compton profile of the target atomic orbitals ${\cal L}_a$, given by
Eq.~(\ref{eq:15}). We compute the Compton profile ${\cal L}_a$ by first solving the Dirac-Fock
equation for the neutral atomic target and then performing a numerical Fourier transform of these
Dirac-Fock orbitals. The numerical Fourier transform was computed by using the routines for the
sine and cosine transforms from NAG library. Finally, the momentum integration in Eq.~(\ref{eq:15})
was evaluated by using the Gauss-Legendre quadratures. The obtained Compton profile ${\cal
L}_a(q_z)$ was stored on a $q_z$ grid and then interpolated to allow a smooth convolution.

\section{Results and discussion}
\label{sec:3}

We start with examining the angle-differential cross section, which is obtained in the laboratory
frame by integrating the double-differential cross section (\ref{eq:21}) over all energetically
possible positron energies,
\begin{align} \label{eq:30}
\frac{d\sigma}{d\Omega_f} = \int d\vare_f\, \frac{d^2\sigma}{d\Omega_f\,d\vare_f}\,.
\end{align}

\begin{figure*}
\centerline{
\resizebox{0.9\textwidth}{!}{%
  \includegraphics{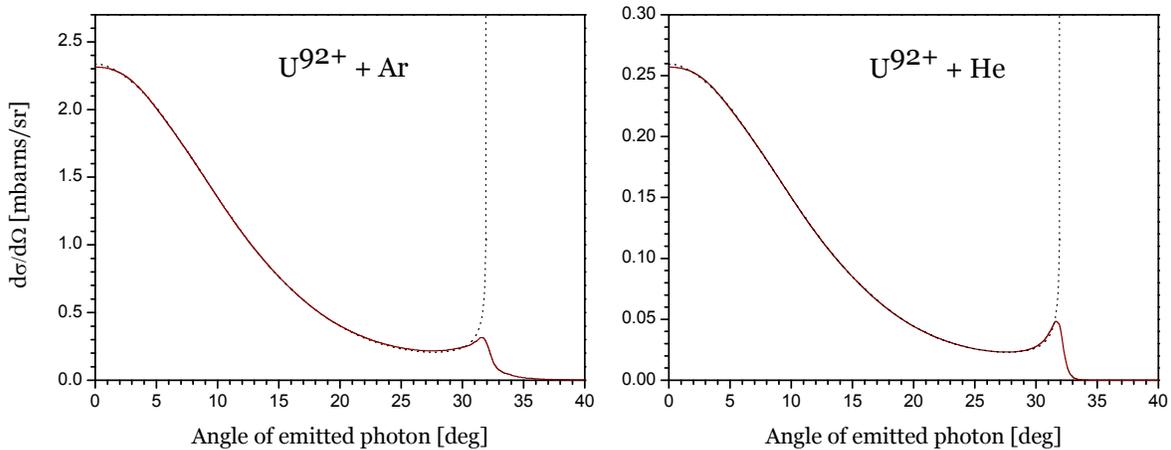}
}}
 \caption{(Color online)
Angle-differential cross section of the NCDR process as a function of the emission angle of
the outgoing positrons in
the laboratory frame. Results are shown for the scattering of U$^{92+}$ on Ar (left graph) and He (right graph)
with capture into the $(1s)^2$ final state of U$^{90+}$, for the projectile energy of 2.37~GeV/u.
The solid line denotes the
results of the full calculation including the Compton profile [Eq.~(\ref{eq:30})],
whereas the dotted line displays data as obtained for the initially monochromatic incoming
electrons.
\label{fig:dcs:1.3}}
\end{figure*}

Let us first consider the dominant NCDR channel, with capture into the ground $(1s)^2$ state of the
final ion. In Fig.~\ref{fig:dcs:1.3} we present a plot of the angle-differential NCDR cross section
as defined by Eq.~(\ref{eq:30}) in the laboratory frame for the collision of bare uranium U$^{92+}$
at 2.37~GeV/u with two neutral targets, Ar (left graph) and He (right graph) and for the capture
into the $(1s)^2$ final state of U$^{90+}$. As the cross sections for different targets are roughly
proportional to the number of target electrons,  both graphs looks quite similar, apart from the
scaling factor of $18/2 = 9$.

In Fig.~\ref{fig:dcs:1.3}, we compare the results that include the momentum distribution of the
electrons in the target (solid line) with the results obtained under the assumption that the
initial electron is free and has a fixed asymptotic momentum and energy $\vare_i = \gamma mc^2$
(dashed line). We note that in the latter case the cross section needs to be multiplied by the
number of electrons in the target, to be directly comparable to the results of the full
calculation. As seen from the figure, the agreement between the two approaches is very good except
in the region close to the critical angle $\theta_{\rm max}$. If the initial electron has fixed
energy and momentum, the laboratory angle cannot exceed $\theta_{\rm max}$ and the differential
cross section is divergent at this point. If we take into account the momentum distribution of the
target electrons, however, the singularity of the cross section is replaced by a distorted
bell-shape of the Compton profile. In addition, emission to the region of $\theta_f > \theta_{\rm
max}$ becomes possible, although the cross section is small in this region and decreases fast as
the angle is increased. As expected, we find a close relation between the width of the Compton
profile of the target and the relative value of the maximum of the peak at $\theta_f = \theta_{\rm
max}$: the sharper the profile, the higher the peak in the differential cross section.

We observe that the enhancement of the cross section in the angular region around $\theta_f =
\theta_{\rm max}$ is not very pronounced and that the maximum of the cross section is reached at
the forward angle $\theta_f = 0^{\circ}$ of the laboratory frame. This makes the forward emission
to be most suitable for determining the NCDR process experimentally. Moreover, the forward angle
seems to be most convenient from the experimental point of view, since the planned setup of the
HESR facility involves a high-resolution forward-emission spectrometer \cite{hillenbrand:15}.

\begin{figure}
\centerline{
\resizebox{0.9\columnwidth}{!}{%
  \includegraphics{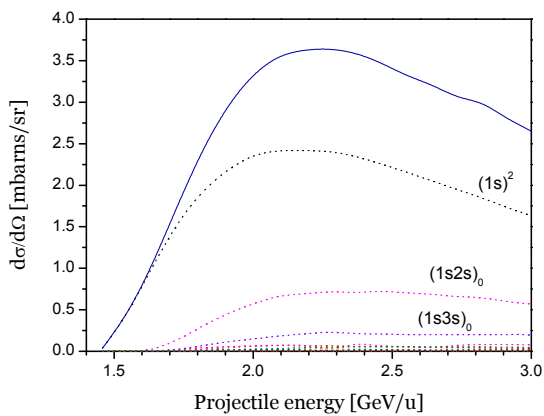}
}}
 \caption{(Color online)
Angle-differential cross section of the NCDR process at fixed emission angle $\theta_f =
0^{\circ}$ of the positrons in the laboratory frame. Theoretical results are shown
as function of the projectile energy, for the collision of
U$^{92+}$ on Ar. The dashed lines denote contributions of capture into different individual final states of
U$^{90+}$, whereas the solid line refers to the sum of all individual contributions.
\label{fig:dcs_energy_000deg}}
\end{figure}

In Fig.~\ref{fig:dcs_energy_000deg} we plot the angle-differential cross section of the NCDR
process for the forward positron emission angle $\theta_f = 0^{\circ}$ as function of the
projectile energy, for capture into different states of the final ion. In the present work, we
performed numerical calculations of the NCDR cross sections with capture into all singly excited
states of the form $(1snl_j)_J$ with $n\le 4$ and double excited states $(2s2l_j)_J$, altogether 36
different NCDR channels. Following to Ref.~\cite{artemyev:09:ncdr}, we observe that the dominant
contribution comes from the capture into the ground $(1s)^2$ state. The second and third largest
contributions come from the capture into the $(1s2s)_0$ and $(1s3s)_0$ states; all other channels
give rise to much smaller contributions.

The energy dependence of the forward-emission NCDR cross section is qualitatively the same as for
the total NCDR cross section. It rapidly increases above the threshold energy, has the flat maximum
at about 2.25~GeV/u and later falls off slowly if the projectile energy is further enlarged. From
this we conclude that the region of projectile energies 2-2.5~GeV/u is the most suitable for
experimental determination of the NCDR process.

We now turn to examining the NCDR spectrum of the double differential cross section (\ref{eq:21})
as function of the positron energy. From the discussion above, we may anticipate that, for a given
positron emission angle, the positron-energy NCDR spectrum has peaks, whose centroid energies are
determined by the binding energy of the final ion state and whose widths are shaped by the Compton
profile of the target atom. In addition, we may expect that in the laboratory frame, capture into
each final state is represented by {\it two} peaks of the positron-energy spectrum.

\begin{figure*}
\centerline{
\resizebox{0.9\textwidth}{!}{%
  \includegraphics{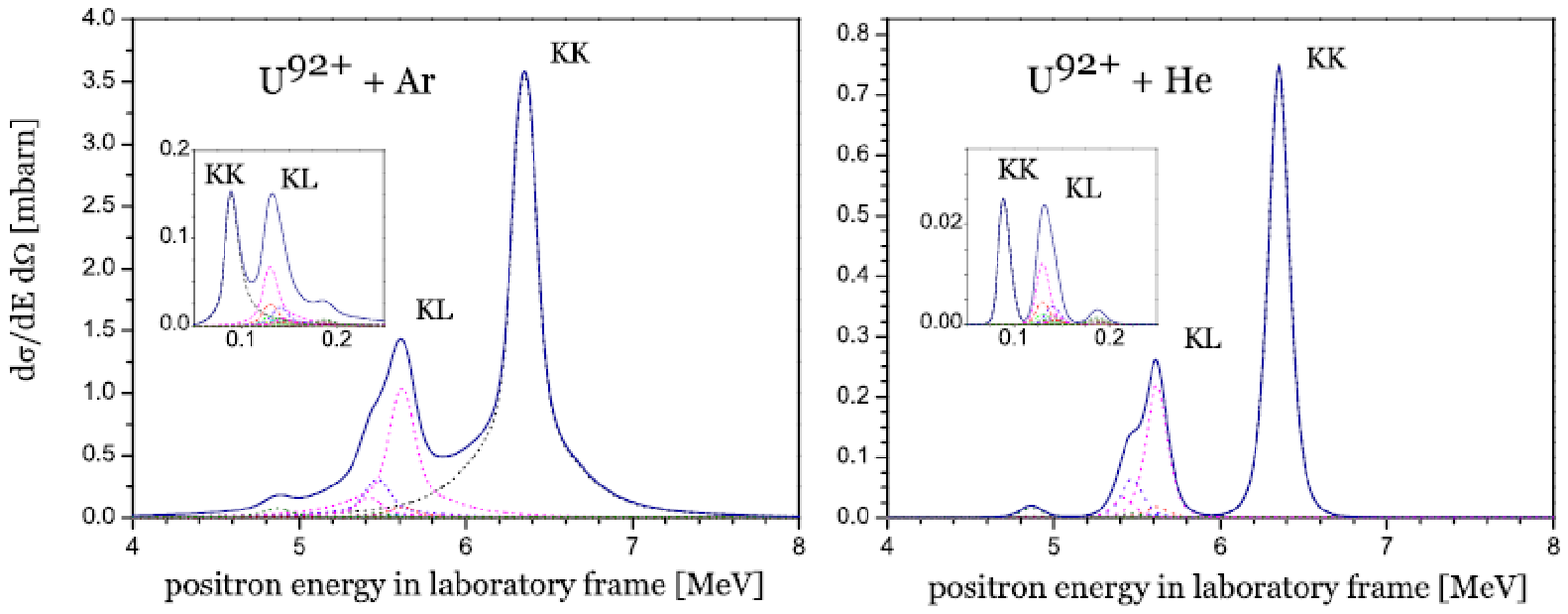}
}}
 \caption{(Color online)
Double-differential cross section of the NCDR process as function of the positron energy
in the laboratory system, for the positron emission angle fixed of $\theta_f = 0^{\circ}$.
Results are shown
for the scattering of U$^{92+}$ at 2.37~GeV/u on Ar (left graph) and He (right graph).
The solid line denotes the total cross section, whereas the dotted lines denote
individual contributions of capture into different final states  of U$^{90+}$.
\label{fig:000deg:1.3}}
\end{figure*}

Fig.~\ref{fig:000deg:1.3} shows the positron-energy spectrum of the NCDR cross section in the
laboratory frame for the forward positron emission angle $\theta_f = 0^{\circ}$, for scattering of
U$^{92+}$ at 2.37~GeV/u on two atomic targets, Ar and He. The dominant NCDR channel of capture into
the ground $(1s)^2$ shell is labelled as ``KK''. As expected, we find two ``KK'' peaks in the
spectrum. The most prominent peak at about $6.4$~MeV corresponds to the forward positron emission
in the projectile frame. The second ``KK'' peak corresponding to the positron backward emission in
the projectile frame is much weaker and is suppressed by more than an order of magnitude. The
individual $(1s2l_j)_J$ peaks in the positron-energy NCDR spectrum cannot be resolved even in the
case of the helium target; the corresponding total peak is labelled as ``KL''. The contributions of
higher excited final states to the positron-energy spectrum are even smaller and are almost
indiscernible on the graphs.

\begin{figure*}
\centerline{
\resizebox{0.9\textwidth}{!}{%
  \includegraphics{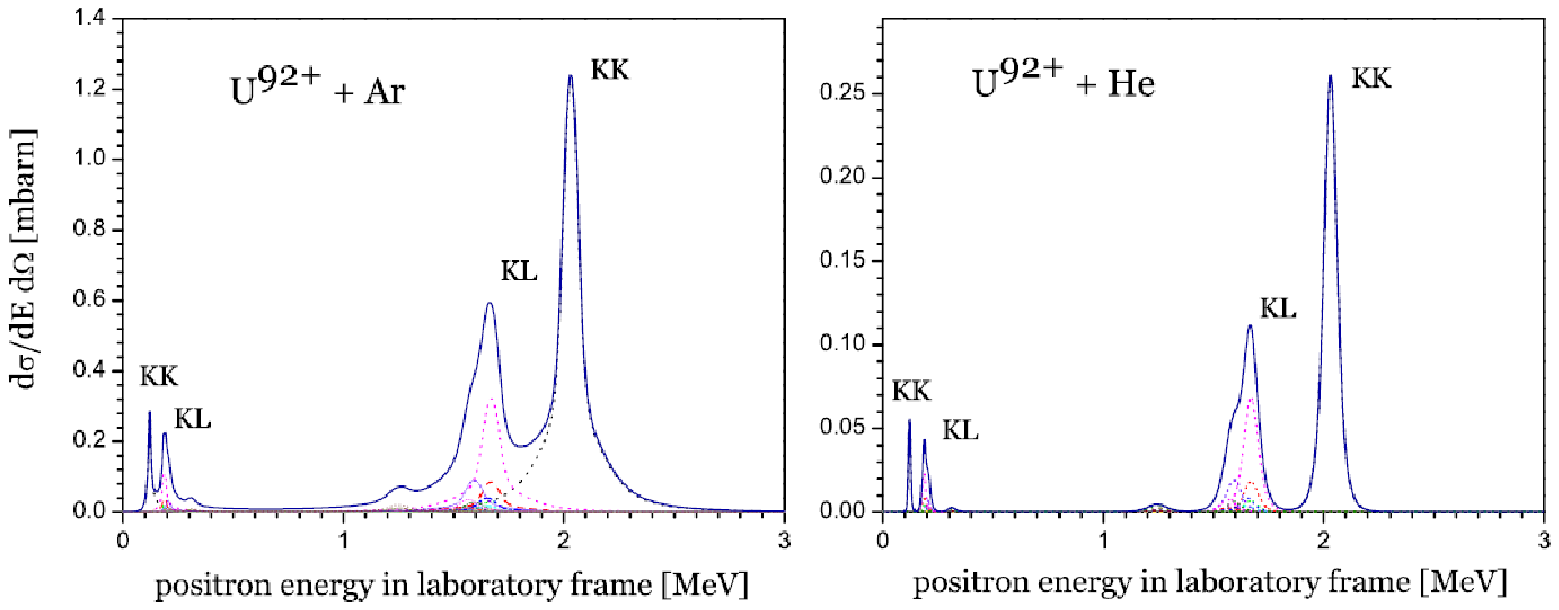}
}}
 \caption{(Color online)
The same as Fig.~\ref{fig:000deg:1.3} but for
$\theta_f = 20^{\circ}$.
\label{fig:020deg:1.3}}
\end{figure*}

The shape of the positron-energy NCDR spectrum changes if the observation angle in the laboratory
frame is increased. Fig.~\ref{fig:020deg:1.3} shows the positron-energy NCDR spectrum for the
positron emission angle $\theta_f = 20^{\circ}$. We observe that the forward- and
backward-scattering peaks are now much closer to each other and that the ratio of intensities of
these peaks is now much (by about 5 times) smaller.

\begin{figure*}
\centerline{
\resizebox{0.9\textwidth}{!}{%
  \includegraphics{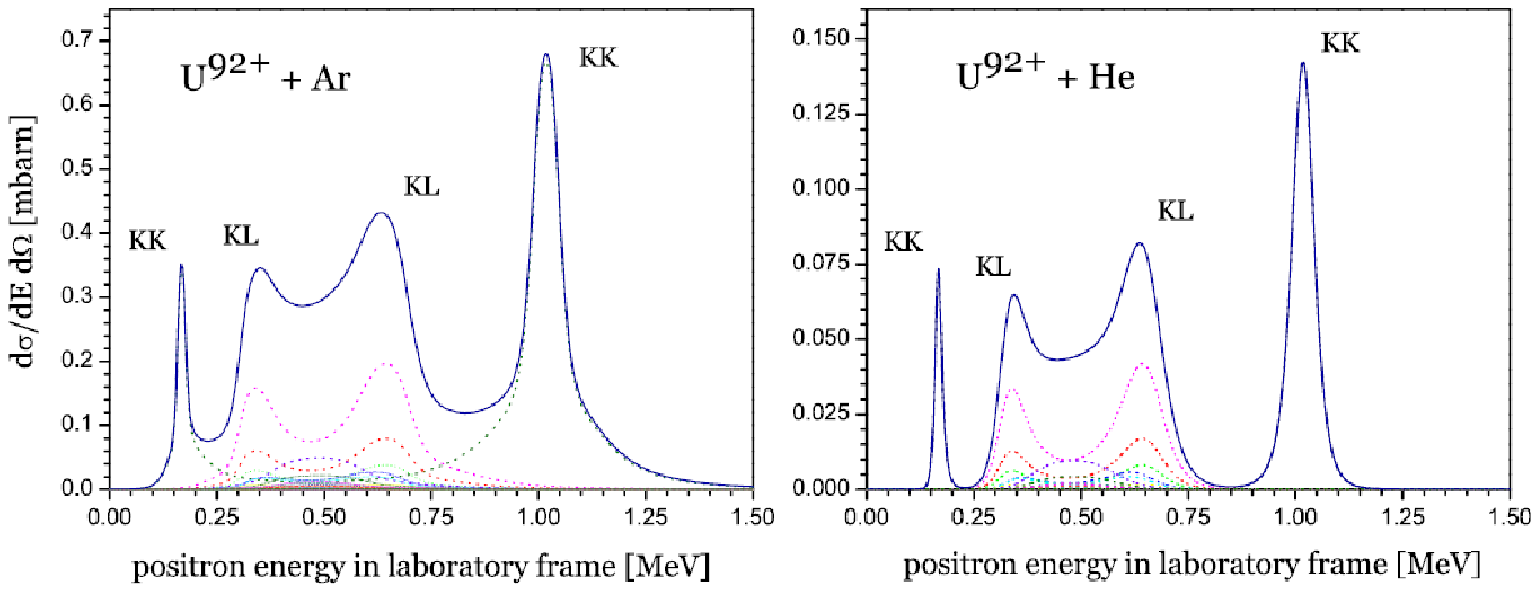}
}}
 \caption{(Color online)
The same as Fig.~\ref{fig:000deg:1.3} but for
$\theta_f = 27^{\circ}$.
\label{fig:027deg:1.3}}
\end{figure*}

The change of the shape of the spectrum becomes even more pronounced if we increase the observation
angle further. Fig.~\ref{fig:027deg:1.3} shows the positron-energy NCDR spectrum for the positron
emission angle $\theta_f = 27^{\circ}$. We see that the two ``KL'' peaks merge into one
double-humped peak and that the two ``KK'' peaks move very close to each other and to the ``KL''
double-peak. For such a large angle, the intensities of the forward- and backward-scattering peaks
become almost equal.

\section{Summary}
\label{sec:4}

In summary, we have studied the process of the negative-continuum assisted dielectronic
recombination (NCDR) that occurs in collision of a heavy bare nucleus with light target atoms. The
single and double-differential cross sections of the positron production in the NCDR process have
been calculated within the  relativistic framework based on QED theory. Different channels of the
NCDR process have been considered explicitly, including the capture into 36 low-lying states of the
final ion. Special attention has been paid to the effects of the atomic targets upon the spectra of
the emitted positrons. It has been demonstrated, in particular, that the target effects remove the
non-physical singularity of the differential cross section, which occurs for the initially
monochromatic electrons.

\section*{Acknowledgements}
This work is supported by BMBF project 05P15SJFAA. V.A.Y. acknowledges support by the Russian
Federation program for organizing and carrying out scientific investigations.




\end{document}